\documentclass[lettersize,journal]{IEEEtran}
\usepackage{amsmath,amsfonts}

\usepackage[algo2e,ruled,vlined,linesnumbered]{algorithm2e}
\SetInd{0em}{0.5em}
\SetKwComment{Comment}{/\!\!/}{}

\usepackage{array}
\usepackage{textcomp}
\usepackage{stfloats}
\usepackage{url}
\usepackage{verbatim}
\usepackage{graphicx}
\usepackage{cite}
\usepackage{subcaption}
\usepackage{format} 
\usepackage{hyperref}
\hypersetup{
    colorlinks = false,
    hidelinks
    }

\usepackage{pgf}
\usepackage{pgfplots}
\usepackage{tikz}
\usetikzlibrary{spy}
\pgfplotsset{compat=1.18}

\usepackage{svg}

\newcommand{\soa}{state-of-the-art }
\newcommand{\WINSZ}{W}
\newcommand{\WINSTRIDE}{s_w}
\newcommand{\sigmatrain}{\sigma^{\mathrm{train}}}
\newcommand{\sigmatest}{\sigma^{\mathrm{test}}}
\newcommand{\Wtrain}{W^{\mathrm{train}}}
\newcommand{\Wtest}{W^{\mathrm{test}}}

\newcommand{\ADJMAT}{\boldsymbol{\Gamma}}
\newcommand{\SIMMAT}{\bm{S}}
\newcommand{\IDMAT}{\bm{I}}

\begin{document}
\title{Fast and Interpretable Nonlocal Neural Networks for Image Denoising via Group-Sparse Convolutional Dictionary Learning}

\author{Nikola Janju\v{s}evi\'{c}, Amirhossein Khalilian-Gourtani, Adeen Flinker, Yao Wang 
\thanks{N. Janju\v{s}evi\'{c} and Y. Wang are with New York University, Electrical and Computer Engineering Department, Brooklyn, NY 11201, USA.}
\thanks{A. Khalilian-Gourtani and A. Flinker are with New York University, Neurology Department, New York, NY 10016, USA.}
\thanks{Please send all correspondence regarding to this manuscript to N. Janju\v{s}evi\'{c} (email:nikola@nyu.edu).}}

\markboth{Submitted to IEEE Transactions on Image Processing, June~2023}%
{Shell \MakeLowercase{\textit{et al.}}: A Sample Article Using IEEEtran.cls for IEEE Journals}

\maketitle

\begin{abstract}
    Nonlocal self-similarity within natural images has become an
    increasingly popular prior in deep-learning models. Despite their successful
    image restoration performance, such models remain largely
    uninterpretable due to their black-box construction. 
    Our previous studies have shown that interpretable
    construction of a fully convolutional denoiser (CDLNet), with performance on par with \soa
    black-box counterparts, is achievable by unrolling a dictionary learning algorithm.
    In this manuscript, we seek an interpretable construction of a convolutional
    network with a nonlocal self-similarity prior that performs on par with
    black-box nonlocal models. We show that such an architecture can be
    effectively achieved by upgrading the $\ell_1$ sparsity prior of 
    CDLNet to a weighted group-sparsity prior. From this
    formulation, we propose a novel sliding-window nonlocal operation, enabled
    by sparse array arithmetic. In addition to competitive performance with black-box
    nonlocal DNNs, we demonstrate the proposed sliding-window sparse attention
    enables inference speeds greater than an order of magnitude faster than its competitors.
\end{abstract}


\section{Background and Introduction}
\IEEEPARstart{N}{onlocal} self-similarity (NLSS) of natural images has proven to
be a powerful signal prior for classical and deep-learning based image restoration.
However, \soa NLSS deep-learning methods are widely constructed as black-boxes, 
often rendering their analysis and improvement beholden to trial and error.
Additionally, current implementations of the NLSS prior in deep-learning separately process 
overlapping image windows, falsely neglecting the dependency between these
overlaps. Here, we address these two shortcomings of nonlocal deep
neural networks (DNNs) from the perspective of interpretable architecture design
and sparse array arithmetic.

A growing literature of DNNs, derived as direct parameterizations of classical
image restoration algorithms, perform on par with \soa black-box fully
convolutional neural networks, without employing common deep-learning tricks
(such as batch-normalization, residual learning, and feature domain processing).
This interpretable construction has been shown to be instrumental in obtaining
parameter and dataset efficiency\cite{janjusevicCDLNet2022, lecouat2020nonlocal, janjusevicGDLNet2022,
Simon2019, Scetbon2019DeepKD}. Our previous work, CDLNet
\cite{janjusevicCDLNet2022}, introduced a unique interpretable construction,
based on convolutional dictionary learning, and achieved novel robustness to
mismatches in observed noise-level during training and inference. By incorporating the NLSS prior into
CDLNet, we demonstrate the first instance of an interpretable network bridging
the performance gap to \soa nonlocal black-box methods for image denoising. 

Nonlocal layers attempt to model long-range image dependencies by computing
pixel-wise self-similarity metrics. To tame the quadratic computational
complexity of this operation, image-restoration DNNs generally rely on computing
an overlapping window NLSS (OW-NLSS) of the input, by which overlapping image windows are
processed independently by the layer/network and subsequently averaged on
overlapping regions to produce the final output \cite{lecouat2020nonlocal, liu2018non}.
Naturally, OW-NLSS incurs a runtime penalty by redundant overlap processing and
a restoration penalty due to the disregard for the correlation among
these overlapping regions. In this work, we propose a novel sliding-window NLSS
(SW-NLSS) operation that addresses these shortcomings.

Previous work used a patch-based group-sparsity prior and OW-NLSS in an
interpretably constructed DNN \cite{lecouat2020nonlocal}. However, this approach
did not achieve competitive performance with black-box NLSS DNNs. In contrast,
we propose to enforce pixel-wise group-sparsity of a latent representation with a dimensionality reduction on the number of channels. We also propose the novel SW-NLSS operation, and achieve denoising
performance on par with the \soa methods at a fraction of the inference time. 

We highlight the following contributions:
\begin{itemize}
    \item a novel and efficient sliding-window nonlocal self-similarity
        operation which addresses the modeling and computational shortcomings of overlapping-window NLSS.
    \item a novel thresholding operation, inspired by a group-sparsity prior, which utilizes a reduced channel dimension of the latent space to achieve \soa inference speeds.
    \item an interpretable nonlocal CNN with competitive
        natural image denoising performance to \soa black-box models.
    \item a fast and open-source implementation\footnote{ 
        \href{https://github.com/nikopj/GroupCDL-TIP}{https://github.com/nikopj/GroupCDL-TIP}.
    } in the Julia programming language \cite{julia}.
\end{itemize}

In Section \ref{sec:prelim}, we introduce the mathematics and notation behind
classical convolutional dictionary learning and group-sparse representation. We
also provide context for related black-box and interpretable deep-learning
methods. In Section \ref{sec:method}, we introduce our sliding-window nonlocal
CNN derived from group-sparse convolutional dictionary learning, dubbed GroupCDL.
In Section \ref{sec:results}, we show experimental results that compare GroupCDL
to \soa deep learning methods.

\section{Preliminaries and Related Work} \label{sec:prelim}
\begin{table}[tb]
\caption{Notation}
\centering
\resizebox{\linewidth}{!}{%
\begin{tabular}{|l|l|} \hline
\multirow{2}{*}{$\x \in \R^{NC}$} & a vector valued image with $N=N_1\times N_2$ pixels, 
\\ & and vectorized channels, $\x =[ \x_1^T,\, \cdots,\, \x_C^T ]^T$. \\
\hline
$\x_c \in \R^N$ & the $c$-th subband/feature-map/channel of $\x$. \\
\hline
$\x[n] \in \R^C$ & the $n$-th pixel of $\x$, $n \in [1, \, N]$. \\
\hline
$\x_c[n] \in \R$ & the $n$-th pixel of the $c$-th channel of $\x$. \\
\hline
$\vec{n}\in [1, N_1] \times [1,N_2]$ & the spatial coordinates of the $n$-th pixel of $\x$. \\
\hline
$\bu \circ \bv \in \R^N$ & the element-wise product of two
vectors. \\
\hline
\multirow{2}{*}{$\bD \in \R^{NC \times QM}$} & a 2D $M$ to $C$
channel synthesis convolution \\ & operator with stride $s_c$, where $Q=N/s_c^2$.\\
\hline
\multirow{2}{*}{$\bD^T \in \R^{QM \times NC}$} & a 2D $C$ to $M$
channel analysis convolution \\ & operator with stride $s_c$, where $Q=N/s_c^2$.\\
\hline
$\bU \in \R^{Q\times N}$ & a $Q \times N$ matrix with elements $\bU_{ij} \in \R$. \\
\hline
$\bU_{i:} \in \R^N, ~ \bU_{:j} \in \R^Q$ & the $i$-th row, $j$-th column of matrix $\bU$. \\
\hline
\multirow{2}{*}{$\bU \otimes \bV \in \R^{QM \times NC}$} & Kronecker product of $\bU \in
\R^{Q\times N}$ and $\bV \in \R^{M\times C}$, \\
& i.e. the block matrix with $\bV$s scaled by $\bU_{ij} ~ \forall ~ i,j$.\\
\hline
$\IDMAT_N \in \R^{N \times N}$ & the $N$ by $N$ identity matrix. \\
\hline
\multirow{2}{*}{$\y = (\IDMAT_C \otimes \bU)\x$} & the matrix $\bU$ applied channel-wise, \\
& i.e. $\y_c = \bU\x_c \in \R^Q ~ \forall ~ 1 \leq c \leq C$. \\
\hline
\multirow{2}{*}{$\y = \overline{\bV}x \equiv (\bV \otimes \IDMAT_N)\x$} & the matrix $\bV$ applied pixel-wise,\\
& i.e. $\y[i] = \bV\x[i] \in \R^M, ~ \forall ~ 1 \leq i \leq N$. \\
\hline
\end{tabular}}
\label{tab:notation}
\end{table}
\subsection{Dictionary Learning and Group-Sparse Representation}
We consider the observation model of additive white Gaussian-noise (AWGN),
\begin{equation} \label{eq:awgn}
    \y = \x + \boldsymbol{\nu}, \quad \textnormal{where}\quad \boldsymbol{\nu}\sim\N(\boldsymbol{0}, \sigma^2 \IDMAT).
\end{equation}
Here, the ground-truth image $\x \in \R^{NC}$ is contaminated with AWGN of noise-level $\sigma$, resulting in
observed image $\y \in \R^{NC}$. In the cases of grayscale and color images,
we consider $\x, \y$ as vectors in $\R^{N}$ or $\R^{N3}$, respectively. 
For convenience and clarity of notation, we denote images in the vectorized form, and any
linear operation on an image as a matrix vector multiplication (see Table
\ref{tab:notation} for details). In implementation, fast algorithms are used
and these matrices are not actually formed, except when explicitly mentioned. 

We frame our signal-recovery problem in terms of a (given)
$s_c$-strided convolutional dictionary $\bD \in \R^{NC \times QM}$, with $Q=N/s_c^2$, i.e. the
columns of $\bD$ are
formed by integer translates of a set of $M$ (vectorized) 2D convolutional
filters, each having $C$ channels.
We assume $\exists \,
\z \in \R^{QM}\, \st \, \x \approx \bD\z$. 
The rich works of sparse-representation and compressed-sensing
provide guarantees based on assumptions of sparsity in $\z$ and regularity on
the columns of $\bD$ \cite{Mallat}. We refer to $\z$ as our sparse-code,
latent-representation, or subband-representation of $\x$.

A popular classical paradigm for estimating $\x$ from an observed
noisy $\y$ is the Basis Pursuit DeNoising (BPDN) model, 
\begin{equation} \label{eq:bpdn}
\underset{\z}{\mathrm{minimize}} ~ \frac{1}{2}\norm{\y -\bD\z}_2^2 + \lambda \psi(\z),
\end{equation}
where $\psi : \R^{QM} \rightarrow \R_+$ is a chosen regularization function. The
Lagrange-multiplier term $\lambda > 0$ provides a trade-off between satisfying
observation consistency and obeying the prior-knowledge encoded by $\psi$. A
popular approach to solving \eqref{eq:bpdn} is the proximal-gradient method
(PGM) \cite{Beck2009}, involving the {\it proximal-operator} of $\psi$, defined as  
\begin{equation} \label{eq:prox}
    \prox_{\tau \psi}(\bv) \coloneqq \argmin_{\x} \tau \psi(\x) +
    \frac{1}{2}\norm{\x - \bv}_2^2, \quad \tau > 0.
\end{equation}
PGM can be understood as a fixed point iteration involving the iterative application
of a gradient-descent step on the $\ell_2$ term of \eqref{eq:bpdn} followed by 
application of the proximal operator of $\psi$,
\begin{equation} \label{eq:pgm}
\z^{(k+1)} = \prox_{\tau \psi}(\z^{(k)} - \eta \bD^T(\bD\z^{(k)} - \y)),
\end{equation}
where $\tau =\eta\lambda$,  and $\eta > 0$ is a step-size parameter.

When $\psi$ is the sparsity-promoting $\ell_1$-norm, the proximal operator is
given in closed-form by element-wise soft-thresholding,
\begin{equation} \label{eq:ST}
\ST_\tau(\z) = \z \circ \left(1 - \frac{\tau}{\abs{\z}}\right)_+,
\end{equation}
where $(\cdot)_+$ denotes projection onto the positive orthant $\R_+$. 
The resulting PGM iterations are commonly referred
to as the Iterative Soft-Thresholding Algorithm (ISTA) \cite{Beck2009}.

More sophisticated priors ($\psi$) can be used to
obtain better estimates of our desired ground-truth image by exploiting
correlations between ``related" image-pixels. One such prior is
{\it group-sparsity}, 
\begin{align} \label{eq:group_sparse}
    \psi(\z) &= \sum_{\substack{m=1 \\j=1}}^{M, Q}
    \sqrt{\sum_{i=1}^Q \ADJMAT_{ij} \z_m[i]^2} 
             = \norm{ \sqrt{(\IDMAT_M \otimes \ADJMAT) \z^2} }_1, 
\end{align}
where $\ADJMAT \in \R_+^{Q\times Q}$ is a row-normalized adjacency matrix (i.e.
$\norm{\ADJMAT_{i:}}_1 = 1$), and $\cdot^2$ and $\sqrt{\cdot}$ are taken element-wise.
Group-sparse regularization may be understood as
encouraging similar latent-pixels to share the same channel-wise sparsity pattern, and has
been shown to improve denoising performance under classical patch-based sparse coding
methods \cite{mairal2009non}, as well as recent interpretably constructed DNNs
\cite{lecouat2020nonlocal}.

In general, the group-sparse
regularizer \eqref{eq:group_sparse} does not have a closed form proximal-operator. 
Motivated by the operator proposed in
\cite{lecouat2020nonlocal}, we propose an approximate solution,
group-thresholding,
\begin{align} \label{eq:GT}
        \GT_{\tau}(\z; \, \ADJMAT) &= \z \circ \left( 1 - \frac{\tau}
            {\sqrt{(\IDMAT_M \otimes \ADJMAT)\z^2}} \right)_+.
\end{align}
Note that the operator proposed in Lecouat et. al \cite{lecouat2020nonlocal}
is equivalent to \eqref{eq:GT} when the adjacency matrix is row-normalized. This
approximate solution has the desirable property of reducing to element-wise
soft-thresholding \eqref{eq:ST} when $\ADJMAT$ is the identity matrix. 

The BPDN \eqref{eq:bpdn} problem can be made more expressive by opting to learn
an optimal dictionary from a dataset of noisy images $\D = \{\y\}$. We express
the (convolutional) dictionary learning problem as,
\begin{equation} \label{eq:dict_learn}
\underset{\{\z\}, \bD \in \mathcal{C}}{\mathrm{minimize}} ~ 
\sum_{\y \in \mathcal{D}} \frac{1}{2}\norm{\y -\bD\z}_2^2 + \lambda
\psi(\z),
\end{equation}
where constraint set $\mathcal{C} = \{ \bD \, : \, \norm{\bD_{:j}}_2^2 \leq 1 ~
\forall \, j \}$ ensures that the regularization term is not rendered useless by
an arbitrary scaling of latent coefficients. Solving \eqref{eq:dict_learn}
generally involves alternating sparse-pursuit (ex. \eqref{eq:pgm}) and a
dictionary update with fixed sparse-codes (ex. projected gradient descent)
\cite{mairal2009online}. 

\subsection{Unrolled and Dictionary Learning Networks}
\begin{figure*}[thb]
    \centering
    \includegraphics[width=\textwidth]{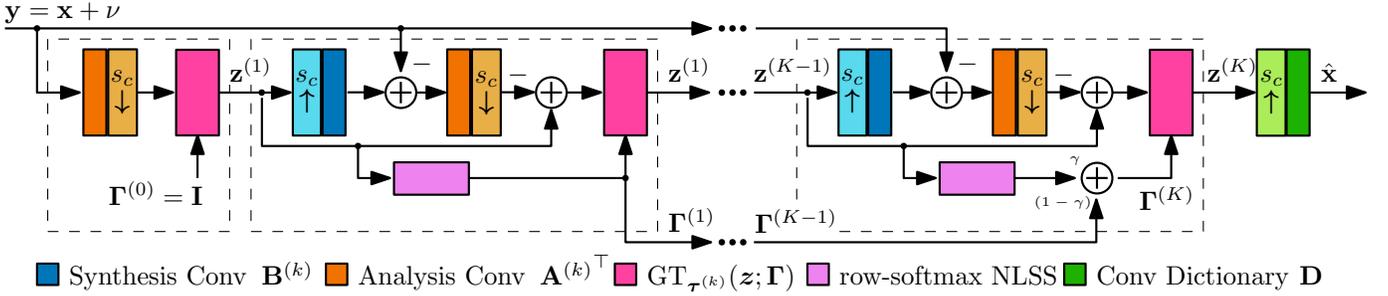}
    \caption{The GroupCDL Architecture. The network begins with no prior of group-sparsity ($\ADJMAT^{(0)} = \bm{I}$). In the second layer, and each subsequent $\Delta K$ layers, the adjacency matrix $\ADJMAT^{(k)}$ is updated by a row-normalized NLSS computation on the latent representation $\z^{(k)}$. NLSS is computed with dense arithmetic (on image patches) during training, and with sparse arithmetic (on the entire image) during inference.}
    \label{fig:arch}
\end{figure*}
 
Approaches in \cite{ongie2020deep, Gilton2019, deqWillet2021} explore the
construction of DNNs as unrolled proximal gradient descent machines with
proximal-operators that are implemented by a black-box CNN, learned end-to-end. 
Although these methods contribute to more principled DNN architecture design in image-processing, their use of black-box neural networks, such as UNets
\cite{unet} and ResNets \cite{he2016deep}, ultimately side-step the goal of full interpretability. In contrast, our previous work CDLNet \cite{janjusevicCDLNet2022} introduces a CNN as a direct parameterization of convolutional PGM \eqref{eq:pgm} with an $\ell_1$ prior, with layers defined as,
\begin{equation} \label{eq:CDLNet}
\begin{gathered}
\z^{(0)} = \bm{0}, \quad \text{for } ~ k=0,1,\dots, K-1,\\
\z^{(k+1)} = \ST_{\boldsymbol{\tau}^{(k)}}\left(\z^{(k)} - {\bm{A}^{(k)}}^T(\bm{B}^{(k)}\z^{(k)} - \y) \right), \\
\btau^{(k)} = \btau^{(k)}_0 + \hat{\sigma}\btau^{(k)}_1, \quad \hat{\x} = \bm{D}\z^{(K)}.
\end{gathered}
\end{equation}
Parameters $\Theta = \{\bD,\, \{{\bA^T}^{(k)}, \,\bB^{(k)}, \, \btau_0^{(k)}, \btau^{(k)}_1 \}_{0=1}^{K-1}\}$ are optimized by back-propagation of a supervised or unsupervised loss function. 
In this manuscript, we extend the formulation and direct parameterization of CDLNet by introducing a novel implementation of the group-sparsity prior, embodied in the proposed GroupCDL architecture (see Section \ref{sec:method}). We also show that the noise-adaptive thresholding of CDLNet, derived from BPDN \eqref{eq:bpdn}, extends to GroupCDL.

Zheng et. al \cite{Zheng_2021_CVPR} propose a DNN architecture based on a
classical dictionary learning formulation of image denoising. However, this network heavily employs black-box models such as UNets
\cite{unet} and multi-layer perceptrons (MLPs). Our proposed method
differentiates itself by using a direct parameterization of variables present
in the classical proximal gradient method \eqref{eq:pgm} with a group-sparsity
regularizer \eqref{eq:group_sparse}. Furthermore, Zheng et. al's
experimental setup does not match a bulk of the existing image denoising literature (by training
on a larger set of images) and their network exists in the ultra-high
parameter count regime ($\approx 17$~M), making a fair comparison beyond the scope
of this paper.

Directly parameterized dictionary learning networks \cite{janjusevicCDLNet2022, janjusevicGDLNet2022, lecouat2020nonlocal, Sreter2018, Simon2019,
Scetbon2019DeepKD} have gained some popularity in recent years due to their
simple design and strong structural similarities to popular ReLU-activation
DNNs. This connection was first established in
the seminal work of Gregor et. al \cite{Gregor2010} in the form of the Learned
Iterative Shrinkage Thresholding Algorithm (LISTA). Here,
we build upon this literature by proposing a novel nonlocal operator \eqref{eq:GTlearned} for the
convolutional dictionary learning formulation of a DNN, derived from a group-sparsity prior \eqref{eq:group_sparse}. 
We also demonstrate that such a network can compete well with (and sometimes outperform)
\soa methods, without sacrificing interpretability. 

Lecouat et. al \cite{lecouat2020nonlocal} propose a nonlocal CNN derived from a
patch-based dictionary learning algorithm with a group-sparsity prior, dubbed
GroupSC. It is well established that the independent processing of image patches
and subsequent overlap and averaging (path-processing) is inherently suboptimal
to the convolutional model, due to the lack of consensus between pixels in
overlapping patch regions \cite{Simon2019}. Our method is in-part inspired by
GroupSC, but proposes a novel version of the group-sparsity mechanism adapted to
the convolutional model and fast application of the network at inference.

\subsection{Nonlocal Networks}
The nonlocal self-similarity prior in image-restoration DNNs is commonly
formulated with OW-NLSS to manage its quadratic complexity \cite{liu2018non,
lecouat2020nonlocal}. The overlap is especially important to ensure that
artifacts do not occur on local-window boundaries. Despite such networks often
being formulated as CNNs, their window-based inference ultimately diminishes the
powerful shift-invariance prior and increases computational cost due to
additional processing of overlapping regions (see Section
\ref{sec:slidingwindow}). 

To reduce computational burden and correctly account for dependencies between
neighboring local windows, we propose a novel sliding-window NLSS, enabled by
sparse matrix arithmetic. Recent works have proposed other so-called ``sparse
attention" mechanisms, however, they have either not been in the context of
image restoration \cite{child2019sparsetransformer}, not employed a
sliding-window \cite{dao2022flashattention}, or have employed a complicated
hashing algorithm to exploit extremely long-range dependencies
\cite{Mei_2021_CVPR}. 

\section{Proposed Method} \label{sec:method}
We consider the problem of image restoration under the AWGN model
\eqref{eq:awgn}, though, in principle, the methods presented may be adapted to
other degradation models with relative ease. Besides being a fundamental building
block of many inverse-problem approaches, AWGN is a popular and successful model
for camera noise after white-balance and gamma-correction \cite{Khashabi2014}. 

\subsection{The GroupCDL Architecture}
We propose a neural network architecture as a direct parameterization of
PGM \eqref{eq:pgm} on the convolutional BPDN problem with a group-sparsity prior, dubbed GroupCDL. 
The GroupCDL architecture is equivalent to replacing the CDLNet
\cite{janjusevicCDLNet2022} architecture's \eqref{eq:CDLNet}
soft-thresholding with group-thresholding w.r.t a row-normalized
adjacency matrix $\ADJMAT$, as described in Algorithm \ref{alg:GroupCDL} and shown in Figure \ref{fig:arch}.
Here, noise-adaptive thresholds are computed using parameters $\btau_0,\, \btau_1 \in \R^M_+$,
${\bA^T}^{(k)}, \bB^{(k)}$ are 2D ($C$ to $M$ channel, stride-$s_c$) analysis and ($M$ to $C$
channel, stride-$s_c$) synthesis convolutions,
respectively, and $\bD$ is our 2D ($M$ to $C$ channel, stride-$s_c$) synthesis
convolutional dictionary. For an input noisy image $\y \in \R^{NC}$, our latent
representation is thus of the form $\z \in \R^{QM}$, where {$Q=N/s_c^2$}.

The adjacency matrix of the group-sparsity prior ($\ADJMAT \in \R^{Q\times Q}_+$) encodes
similarity between latent subband pixels $\z[i],\, \z[j] ~ \forall \, i,\, j$.
To manage computational complexity while staying true to the
convolutional inductive bias of the network, we form this adjacency using a local
sliding-window of size $\WINSZ \times \WINSZ$.
Motivated by the nonlocal similarity computations of black-box networks
\cite{liu2018non}, we compute the similarity after transforming $\z[i]$ and
$\z[j]$ along the channel dimension. 
Specifically, we first compute similarities at the $k$-th layer of the network as,
\begin{equation} \label{eq:sim}
    \SIMMAT^{(k)}_{ij} = 
    \begin{cases}
        -\norm{\bW_\theta \z^{(k)}[i] - \bW_\phi \z^{(k)}[j]}_2^2, &
        \norm{\vec{i} - \vec{j}}_\infty \leq \WINSZ \\
        - \infty , & \text{otherwise}
    \end{cases} 
\end{equation}
where
$\bW_\theta, \bW_\phi \in \R^{M_h \times M}$ are learned pixel-wise transforms shared across all layers. 
That is, similarities $\SIMMAT^{(k)}_{ij}$ are only computed for spatial locations $i$ and $j$ within a $W\times W$ window centered on $i$.
The similarity matrix is then normalized via a row-wise softmax
operation. To reduce computational complexity, we only compute similarity every
$\Delta K$ layers. 
We employ a convex combination of this normalized similarity and the adjacency
of the previous layer ($\ADJMAT^{(k-1)}$) with a learned
parameter $\gamma \in [0,1]$ (see Alg. \ref{alg:GroupCDL}), to ensure smooth updates.
We consider circular boundary conditions when forming nonlocal windows in
\eqref{eq:sim}, resulting in a block-circulant with circulant block sparsity
pattern for $\SIMMAT$ and $\ADJMAT$, as depicted in Figure \ref{fig:nlss}. 

Mimicking the use of subband transforms in the similarity computation
\eqref{eq:sim}, we introduce two additional subband transforms, 
$\bW_\alpha \in
\R^{M \times M_h}, \bW_\beta \in \R_+^{M \times M_h}$, into the
group-thresholding operation,
\begin{equation} \label{eq:GTlearned}
    \begin{gathered}
    \GT_{\btau}(\z; \, \ADJMAT) = \z \circ
    \left(1 - \frac{\btau}{\bxi} \right)_+, \\
    \bxi = \overline{\bW_\beta} \sqrt{(\IDMAT_{M_h} \otimes \ADJMAT) (\overline{\bW^T_\alpha} \z)^2}, \\
    \end{gathered}
\end{equation}
where $\bxi
\in \R_+^{QM}$ contributes to the image-adaptive spatially varying threshold. Here, $\overline{\bW}$ refers to a pixel-wise application of a matrix $\bW$ (see Table \ref{tab:notation}).
In contrast to \eqref{eq:GT}, $\bW_\alpha$ allows the adjacency-weighted energy of the latent representation to be computed in a compressed subband domain (by setting $M_h << M$). Then, $\bW_\beta$
maps this energy back to the uncompressed subband domain ($M$ channels), pixel-wise.
In Section \ref{sec:ablation}, we empirically show that the use of a compressed
subband domain has a positive impact on denoising performance and an even
greater impact on reducing inference time. 

\begin{algorithm2e}[tb]
\caption{Group-sparse Convolutional Dictionary Learning Network (GroupCDL) Forward Pass}
\label{alg:GroupCDL}
\begin{small}
\textbf{Input:} noisy image $\y$, estimated noise-level $\hat{\sigma}$ \;
\textbf{Parameters:} $\Theta = \{\gamma,\, \bW_{\{\theta, \phi, \alpha, \beta\}}, \, \bD,\, \{{\bA^T}^{(k)}, \,\bB^{(k)}, \, \btau^{(k)}_{\{0,1\}} \}_{0=1}^{K-1}\} $ \;
\textbf{Preprocess:} $\tilde{\y} =\y - \mathrm{mean}(\y)$ \;
\textbf{Initialize:} $\z^{(0)} = \mathbf{0}$, $\ADJMAT^{(0)} = \bm{I}$,  $\btau^{(k)} = \btau^{(k)}_0 + \hat{\sigma}\btau^{(k)}_1 \,\, \forall \,\, k$ \;
\For{$k=0,1,\dots,K-1$}{
    \Comment{Update row-normalized adjacency}
    \If{$k=1$}{
        $\ADJMAT^{(1)} = \softmax(\SIMMAT^{(1)})$ \Comment{Eq. \eqref{eq:sim}}
    }
    \ElseIf{$\mathrm{mod}(k+1,\, \Delta K) = 0$}{
        $\ADJMAT^{(k)} = \gamma\softmax(\SIMMAT^{(k)}) + (1-\gamma)\ADJMAT^{(k-1)}$ \Comment{Eq. \eqref{eq:sim}}
    }
    \Else{
        $\ADJMAT^{(k)} = \ADJMAT^{(k-1)}$ \;
    }
    $\bm{r} = {\bA^T}^{(k)}(\bB^{(k)}\z^{(k)} - \tilde{\y})$\;
    $\z^{(k+1)} = \GT_{\btau^{(k)}}\left(\z^{(k)} - \bm{r}; \ADJMAT^{(k)} \right)$ \Comment{Eq. \eqref{eq:GTlearned}}
}
\textbf{Output:} $\hat{\x} = \bD \z^{(K)} + \mathrm{mean}(\y)$ \;
\end{small}
\end{algorithm2e}
\subsection{Sliding-window vs. Patch based Self-Attention} \label{sec:slidingwindow}
\begin{figure}[thb]
    \centering
    \begin{subfigure}{\linewidth}
        \centering
        \includegraphics[width=\linewidth]{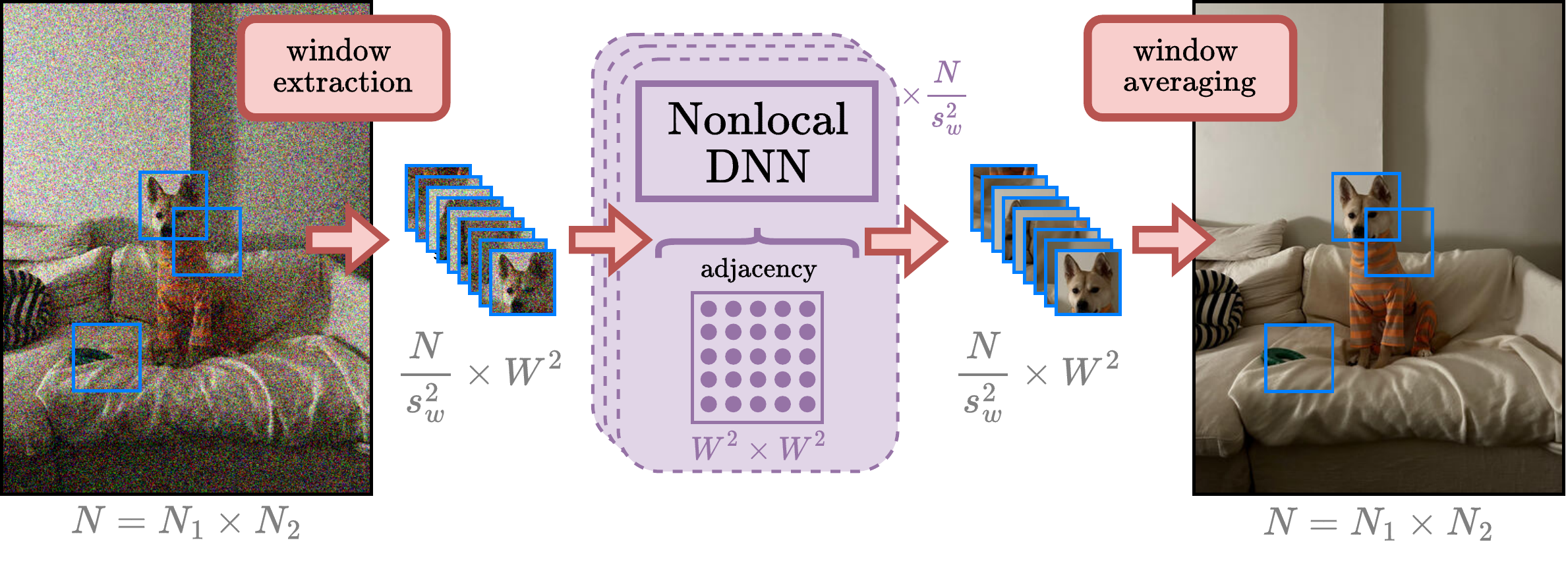}
        \caption{OW-NLSS}
        \vspace*{1em}
    \end{subfigure}
    \begin{subfigure}{\linewidth}
        \centering 
        \includegraphics[width=0.9\linewidth]{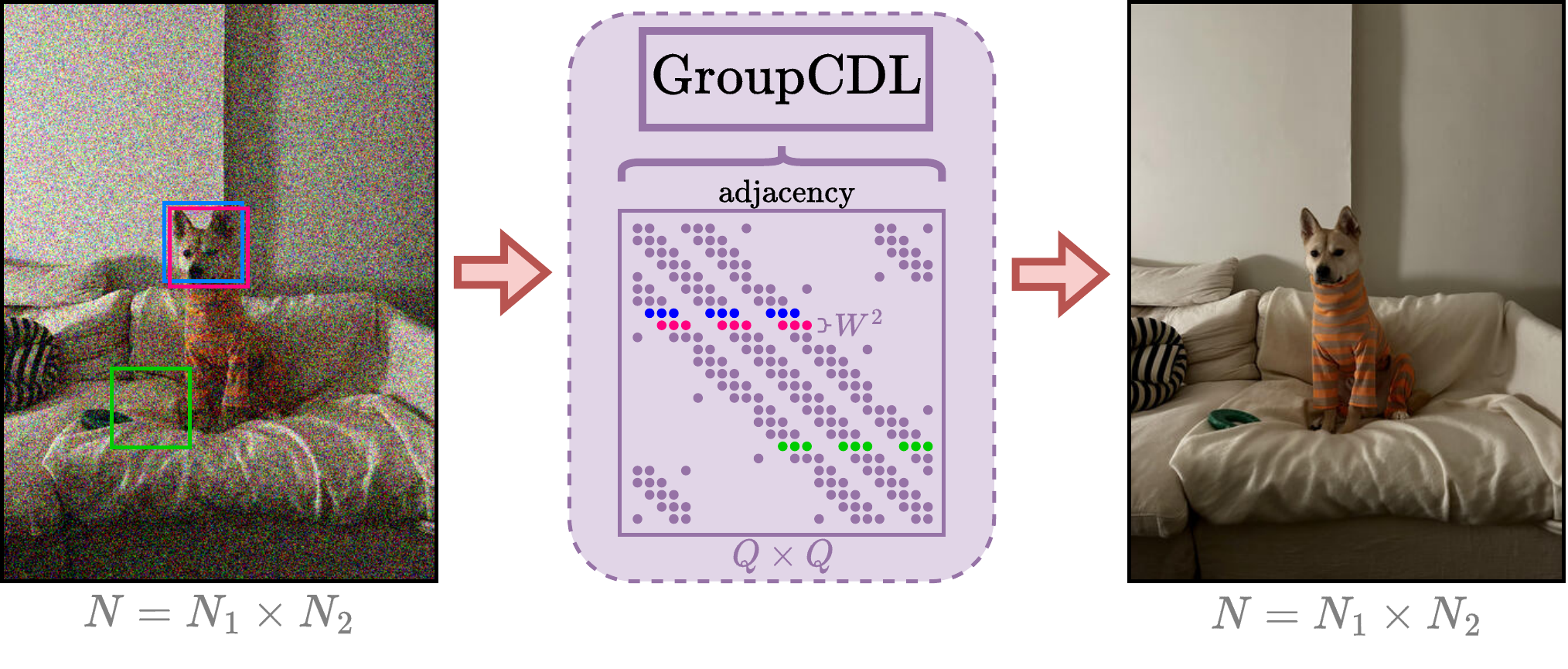}
        \caption{SW-NLSS}
    \end{subfigure}
    \caption{
        (a) In OW-NLSS, the input image is divided into overlapping windows (of size $W\times W$ and with window-stride $s_w \times s_w$), processed
        independently via a DNN with dense self-attention. The denoised windows
        are then placed in their original positions and averaged on their
        overlaps. (b) In the proposed SW-NLSS, the entire image is processed in a single forward pass, made possible by sparse matrix arithmetic. The
        adjacency matrix has a block-circulant with circulant blocks (BCCB)
        sparsity pattern, where the number of non-zeros in each row/column is
        at most $\WINSZ^2$. The adjacency matrix is computed in the subband
        domain, with spatial dimension $Q=N/s_c^2$, where $s_c$ is the
        convolution stride. Hence, the effective image-domain window-size is $s_cW \times s_cW$.
    }
    \label{fig:nlss}
\end{figure}
The SW-NLSS employed by GroupCDL \eqref{eq:sim} is favorable to the independent OW-NLSS employed by GroupSC
\cite{lecouat2020nonlocal} and black-box DNNs \cite{liu2018non, zhang2019residual}, 
because it naturally encourages agreement on overlapping regions and centers
the nonlocal windows on each pixel. 
As shown in Figure \ref{fig:nlss}, OW-NLSS additionally incurs computational
overhead by processing overlapping pixels multiple times. This burden inherent
to OW-NLSS can be expressed in terms of the image dimensions $N = N_1 \times N_2$,
window-size $\WINSZ \times \WINSZ$, and window-stride $\WINSTRIDE \times
\WINSTRIDE$. We express the burden factor as a ratio of the number of pixels
processed by a single OW-NLSS layer over an SW-NLSS layer,
\begin{equation} \label{eq:burden}
    \frac{N_1/\WINSTRIDE \times N_2/\WINSTRIDE \times \WINSZ^2}{N_1 \times N_2} = \frac{\WINSZ^2}{\WINSTRIDE^2}.
\end{equation}
Common nonlocal window sizes, $45 \times 45$,  and window-strides, $7
\times 7$, such as used by NLRN \cite{liu2018non}, make this burden factor 41 times the
computational complexity of an equivalent SW-NLSS. GroupCDL's use of strided
convolution may add an additional $s_c^2 \times$  computational benefit compared
to common NLSS implementations by computing similarities over a reduced spatial
dimension $Q = N/s_c^2$. We further explore the relation between computation
time and denoising performance of these two NLSS inference methods in Section
\ref{sec:exp:slidingwindow}.

\subsection{Group-Thresholding vs. Black-box Attention}
Nonlocal self-similarity is used across domains in DNNs, from transformer
architectures \cite{dao2022flashattention} to nonlocal image restoration networks
\cite{liu2018non, zhang2019residual}. The underlying formula behind these methods is most commonly dot-product attention (DPA), given below,
\begin{align} \label{eq:dotprod_atten_2}
    \z^{(k+1)} &= (I_{M_{\mathrm{out}}} \otimes \ADJMAT) \overline{\bW_v} \z^{(k)} \\
    \ADJMAT &= \softmax(\SIMMAT^{(k)}) \\
    \SIMMAT^{(k)}_{ij} &= \frac{\z^{(k)}[j]^T \bW_q^T \bW_k
    \z^{(k)}[i]}{\sqrt{M_{h}}}, \label{eq:dot_sim}
\end{align}
where $\z^{(k)} \in \R^{NM_{\mathrm{in}}}$, $\bW_q, \bW_k, \in \R^{M_{h} \times M_{\mathrm{in}}}$,
$\bW_v \in \R^{M_{\mathrm{out}} \times M_{\mathrm{in}}}$, and $\z^{(k+1)} \in \R^{NM_{\mathrm{out}}}$. 
Learned matrices $\bW_q, \bW_k, \bW_v$ are understood to transform the input signal
$\z^{(k)}$ to so-called query, key, and value signals. 

Both DPA and the proposed GT \eqref{eq:GTlearned} make use of a normalized adjacency matrix
($\ADJMAT$), computed in an asymmetric feature domain. Both use this adjacency
to weight the current spatial features, identically over channels. However, in
DPA, the weighting directly results in the layer's output (via matrix
multiplication, see \eqref{eq:dotprod_atten_2}), whereas in GT, this weighting informs a spatially adaptive
soft-thresholding. 

The proposed GT's decoupling of adjacency application and output dimension is key in allowing
group-thresholding to be computationally efficient, as the adjacency
matrix-vector multiplication can be performed in a compressed subband domain. In contrast,
DPA operating in a compressed feature domain ($M_{\mathrm{out}} << M_{\mathrm{in}}$) would harm the
capacity of the network's latent representation. In Section
\ref{sec:ablation} we show empirical evidence for favoring the negative-norm
similarity in GT over the dot-product similarity of DPA.

\section{Experimental Results} \label{sec:results}
\subsection{Experimental Setup}
\textbf{Architecture}: We denote the network detailed in Algorithm \ref{alg:GroupCDL} as
GroupCDL. GroupCDL and CDLNet are trained with noise-adaptive thresholds ($\btau^{(k)} = \btau^{(k)}_0 + \hat{\sigma}\btau^{(k)}_1$) unless specified using the
-B suffix, indicating the models are noise-blind ($\btau^{(k)} = \btau^{(k)}_0$). The hyperparameters for these architectures are
given in Table \ref{tab:arch}, unless otherwise specified.

\begin{table}[tb]
\caption{Architectures of the GroupCDL models, CDLNet models, and variants presented in the experimental section. 
We use $C=3$ and $C=1$ for color and grayscale denoising networks, respectively.
A filter size of $7\times 7$ is used for all models. Conv-stride
$s_c=2$, adjacency update period $\Delta K=5$, is used unless otherwise specified.}
\centering
\resizebox{\linewidth}{!}{%
\begin{tabular}{rccccc} \hline
    Name & Task & $K$ & $M$ & $M_h$ & $W$ \\ \hline
     CDLNet(-S,-B) & Gray & 30 & 169 & - & - \\ 
     GroupCDL(-S,-B) & Gray & 30 & 169 & 64 & 35 \\ 
     CDLNet(-S,-B) & Color & 24 & 96 & - & - \\ 
     GroupCDL(-S,-B) & Color & 24 & 96 & 48 & 35 \\
     \hline
\end{tabular}
}
\label{tab:arch}
\end{table}

\textbf{Dataset and Training}:
Let $f_\Theta$ denote the GroupCDL DNN as a function of
parameters $\Theta$. Let $\D = \{(\y, \sigma, \x)\}$ denote a dataset of noisy and
ground-truth natural image pairs, with noise-level $\sigma$. Grayscale (and color) GroupCDL models are trained on
the (C)BSD432 \cite{bsd} dataset with the supervised mean squared
error (MSE) loss,
\begin{equation} \label{eqn:mse}
\underset{
\substack{
\bW_\theta, ~ \bW_\phi,~\bW_\alpha,\\
 \bW_\beta \geq 0,  ~ \gamma \in [0, 1], \\
\bm{D} \in \mathcal{C}, ~ \{ \btau^{(k)} \geq 0 \}_{k=0}^{K-1}, \\
\{ \bm{A}^{(k)} \in \mathcal{C}, ~ \bm{B}^{(k)} \in \mathcal{C} \}_{k=0}^{K-1}
}
}{\mathrm{minimize}} \quad \sum_{\{\y, \sigma, \x\} \in \mathcal{D}} 
\norm{f_{\Theta}(\y, \sigma) - \x}_2^2,
\end{equation}
where $\mathcal{C} = \{ \bD \, : \, \norm{\bD_{:j}}_2^2 \leq 1 ~
\forall \, j \}$. We use the Adam
optimizer with default parameters \cite{adam}, and project the network parameters
onto their constraint sets after each gradient step.
The dataset is generated online with random crops, rotations, flips, and
AWGN of noise-level $\sigma$ sampled uniformly within $\sigmatrain$ for each mini-batch element. All models are trained with the same hyperparameters given in
\cite{janjusevicCDLNet2022}, however, GroupCDL models are trained for 270k
iterations with a batch-size of 32.

Test and validation performance is evaluated on several datasets. The dataset name, along with (arithmetic) average 
dimensions, are provided to better understand reported inference timings: Set12 (362 $\times$ 362), CBSD68 \cite{bsd} (481 $\times$ 321), Urban100 \cite{Urban100} (1030 $\times$ 751), and 
NikoSet10\footnote{see \href{https://github.com/nikopj/GroupCDL-TIP}{https://github.com/nikopj/GroupCDL-TIP}.} (1038 $\times$ 779).

\textbf{Training Initialization}:
CDLNet models are initialized as ISTA with $\btau_0 = 10^{-2}, ~ \btau_1 = 0$, and
a base dictionary $\bD$ that has been spectrally normalized. Details are given
in \cite{janjusevicCDLNet2022}. GroupCDL models are initialized with a trained
CDLNet model. Pixel-wise transforms $\bW_{\{\theta, \phi, \alpha\}}$ are initialized with the same weights drawn from a Glorot Uniform distribution centered at zero \cite{glorot}, and $\bW_\beta$ is drawn from a similar Glorot Uniform distribution, tailored to the positive orthant. We initialize parameter $\gamma = 0.8$.

\textbf{Group-Thresholding Training vs. Inference}:
A GroupCDL model, with nonlocal window-size $\WINSZ$ and conv-stride $s_c$, is
trained using image crops of dimension $s_c \WINSZ \times s_c \WINSZ$ such that a
single nonlocal window is used during training. Dense arithmetic is used in
construction and application of the normalized adjacency matrix $\ADJMAT^{(k)}$
throughout training.

On inference of a noisy image $\y \in \R^{NC}$, with latent representation
$\z \in \R^{QM}$ ($Q=N/s_c^2$), the adjacency matrix $\ADJMAT~\in~\R_+^{Q \times
Q}$ is constructed with a block circulant with circulant blocks (BCCB) sparsity
pattern and sparse matrix arithmetic is used.
The adjacency matrix requires an allocation of 
$Q \times \WINSZ^2 \times \text{\texttt{sizeof(float)}}$ bytes. 
For a high resolution image of size $1024 \times 1024$ (such as images in the
Urban100 dataset \cite{Urban100}) and a conv-stride of $2$, this
results in roughly {$2.6$~GB} of memory required to store $\ADJMAT^{(k)}$ and
$\ADJMAT^{(k-1)}$, which is by and large the memory consumption of inference.

\textbf{Hardware}:
All models were trained on a single core Intel Xeon CPU with 2.90 GHz clock and a single NVIDIA A100 GPU.
CDLNet pre-training takes roughly 6 hours, and GroupCDL training takes
approximately an additional 36 hours. Note that training and inference can take place on GPUs
with as low as 16 GB of memory. For code-base compatibility reasons, all inference timings in Tables
\ref{tab:graysingle}, \ref{tab:colorsingle} 
were determined by running models (provided by the authors of the respective papers), and our trained GroupCDL/CDLNet models, on
a single core Intel Xeon CPU with 2.90 GHz clock and a NVIDIA Quadro RTX-8000
GPU.

\subsection{Single Noise-Level Performance} \label{sec:exp:single}
In this section, we demonstrate the fast inference speed and competitive
denoising performance of the proposed GroupCDL. All models are trained on a single noise-level and tested at the same noise-level ($\sigmatrain = \sigmatest$). We compare GroupCDL to its fully convolutional counterpart (CDLNet), the non-learned nonlocal method BM3D \cite{bm3d, bm3d-gpu}, popular
local and nonlocal black-box CNNs \cite{DnCNN,Valsesia2020,liu2018non}, and a
patch-processing dictionary learning based nonlocal DNN (GroupSC)
\cite{lecouat2020nonlocal}. 
\begin{table*}[ht]
\centering
\caption{Grayscale denoising performance (PSNR (dB)/ $100\times$SSIM) and GPU
inference runtimes. All learned methods are trained on BSD432 \cite{bsd} with an MSE loss function
($\sigma = \sigmatrain = \sigmatest$). Learned parameter counts are displayed
below the method names.}
\resizebox{\linewidth}{!}{%
\begin{tabular}{cccccccccc} \hline
\multirow{2}{*}{Dataset} & \multirow{2}{*}{Noise $\sigma$} & BM3D \cite{bm3d} & DnCNN \cite{DnCNN} &
CDLNet \cite{janjusevicCDLNet2022} & GroupSC \cite{lecouat2020nonlocal} & GCDN
\cite{Valsesia2020} & NLRN \cite{liu2018non} & GroupCDL-S \\
& & - & 556k & 507k & 68k & 6M & 340k & 550k \\\hline
\multirow{3}{*}{Set12} 
& 15 & 32.37/89.52 & 32.86/90.31 & 32.87/90.43 & 32.85/90.63 & \underline{33.14}/\underline{90.72} & {\bf 33.16}/90.70 & 33.05/{\bf 90.73} \\
& 25 & 29.97/85.04 & 30.44/86.22 & 30.52/86.55 & 30.44/86.42 &
\underline{30.78}/86.87 & {\bf 30.80}/\underline{86.89} & 30.75/{\bf 86.93} \\
& 50 & 26.72/76.76 & 27.18/78.29 & 27.42/79.41 & 27.14/77.97 & 27.60/79.57 & {\bf 27.64}/\underline{79.80} & \underline{27.63}/{\bf 80.04} \\
time (s) & & 0.010 & 0.119 & 0.019 & 22.07 & 404.8 & 25.62 & 0.68 \\\hline

\multirow{3}{*}{BDS68 \cite{bsd}} 
         & 15 & 31.07/87.17 & 31.73/89.07 & 31.74/89.18 & 31.70/{\bf 89.63} &
\underline{31.83}/89.33 & {\bf 31.88}/89.32 & 31.82/\underline{89.41} \\
                        & 25 & 28.57/80.13 & 29.23/82.78 & 29.26/83.06 & 29.20/\underline{83.36} & 29.35/83.32 &
{\bf 29.41}/83.31 & \underline{29.38}/{\bf 83.51} \\
& 50 & 25.62/68.64 & 26.23/71.89 & 26.35/72.69 & 26.17/71.83 & 26.38/{\bf 73.89}
& {\bf 26.47}/72.98 & {\bf 26.47}/\underline{73.32} \\
time (s) & & 0.011 & 0.039 & 0.022 & 23.63 & 539.7 & 26.66 & 0.65 \\\hline

\multirow{3}{*}{Urban100 \cite{Urban100}} 
& 15 & 32.35/92.20 & 32.68/92.55 & 32.59/92.85 & 32.72/93.08 & {\bf 33.47}/{\bf 93.58} & \underline{33.42}/\underline{93.48} & 33.07/93.40 \\
& 25 & 29.70/87.77 & 29.92/87.97 & 30.03/89.00 & 30.05/89.12 & {\bf 30.95}/{\bf 90.20}
& \underline{30.88}/\underline{90.03} & 30.61/\underline{90.03} \\
& 50 & 25.95/77.91 & 26.28/78.74 & 26.66/81.11 & 26.43/80.02 & {\bf 27.41}/81.60 & \underline{27.40}/\underline{82.44} & 27.29/{\bf 83.05} \\
time (s) & & 0.030 & 0.096 & 0.090 & 93.33 & 1580 & 135.8 & 3.56 \\\hline
\end{tabular}
}
\label{tab:graysingle}
\end{table*}

\begin{table}[ht]
\centering
\caption{Color Denoising performance, PSNR (dB), and GPU inference runtimes on CBSD68 \cite{bsd} . All learned methods are trained on CBSD432 \cite{bsd} ($\sigma = \sigma^{\mathrm{train}} = \sigma^{\mathrm{test}}$).}
\resizebox{\linewidth}{!}{%
\begin{tabular}{cccccccccc} \hline
    \multirow{2}{*}{Model} & \multirow{2}{*}{Params} & \multirow{2}{*}{time (s)} & \multicolumn{5}{c}{Noise-level ($\sigma$)} \\
                       & & & 10 & 15 & 25 & 30 & 50 \\ \hline
BM3D \cite{bm3d}                   & -    & 0.019 & 34.56 & 33.49 & 30.68 & 28.05 & 27.36 \\
DnCNN \cite{DnCNN}                 & 668k & 0.054 & 36.31 & 33.99 & 31.31 & -     & 28.01 \\
CDLNet                             & 694k & 0.009 & 36.31 & 34.04 & 31.39 & 30.52 & 28.18 \\
GroupSC \cite{lecouat2020nonlocal} & 119k & 40.81 & 36.40 & 34.11 & 31.44 & 30.58 & 28.05 \\
RNAN \cite{zhang2019residual}      & 8.96M& 1.92 & {\bf 36.60} &  -    &  - & {\bf 30.73} & \underline{28.35} \\
GroupCDL-S                         & 698k & 0.39  & \underline{36.43} &
{\bf 34.19} & {\bf 31.58} & \underline{30.70} & {\bf 28.37} \\
\hline
\end{tabular}
}
\label{tab:colorsingle}
\end{table}
\begin{figure}[thb]
    \centering
    \captionsetup{justification=centering,singlelinecheck=false}
    \input{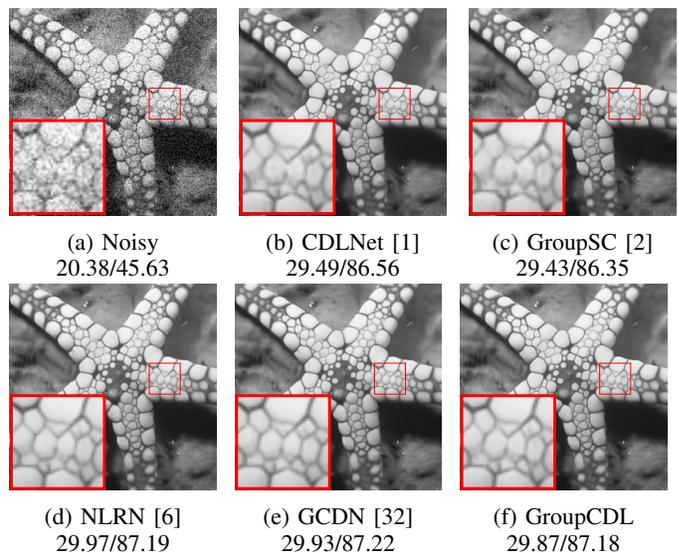}
    \captionsetup{justification=justified,singlelinecheck=false}
    \caption{
        Visual comparison for grayscale denoising models
        ($\sigmatest=\sigmatrain=25$) on the starfish test image. PSNR (dB)/$100 \times$SSIM reported for each image.
    }
    \label{fig:graysingle}
\end{figure}
Table \ref{tab:graysingle} shows the grayscale
denoising performance and inference speed of the aforementioned models across
several common datasets. We
include a learned parameter count as a crude measure of expressivity of the model.
The group-sparsity prior of GroupCDL significantly increases
denoising performance compared to the unstructured sparsity prior of CDLNet, though at a detriment to inference speed. 
We observe that GroupCDL has denoising performance superior to other dictionary
learning based networks using group sparsity prior (GroupSC) and competitive performance with \soa
black-box methods (GCDN, NLRN). Most notably, GroupCDL has the fastest
inference runtime among nonlocal methods, with at least an order of magnitude
difference. These timing differences between GroupCDL and NLRN (or GroupSC)
correspond well with the analysis in Section \ref{sec:slidingwindow} regarding
the use of sliding-window sparse nonlocal processing vs. overlapping-window nonlocal
processing. 

Table \ref{tab:colorsingle} shows the color image denoising performance and
inference speed of the aforementioned classical benchmark \cite{bm3d, bm3d-gpu}, local DNN
\cite{DnCNN}, nonlocal patch-processing DNN \cite{lecouat2020nonlocal}, and black-box nonlocal
DNN \cite{zhang2019residual} against the proposed GroupCDL. We observe that
GroupCDL outperforms the interpretable patch-processing dictionary learning
DNN (GroupSC) and CNNs (DnNN, CDLNet). GroupCDL performs competitively to the black-box nonlocal DNN (RNAN \cite{zhang2019residual}) at a faction of the learned parameter count and inference time.
\begin{figure*}[thb]
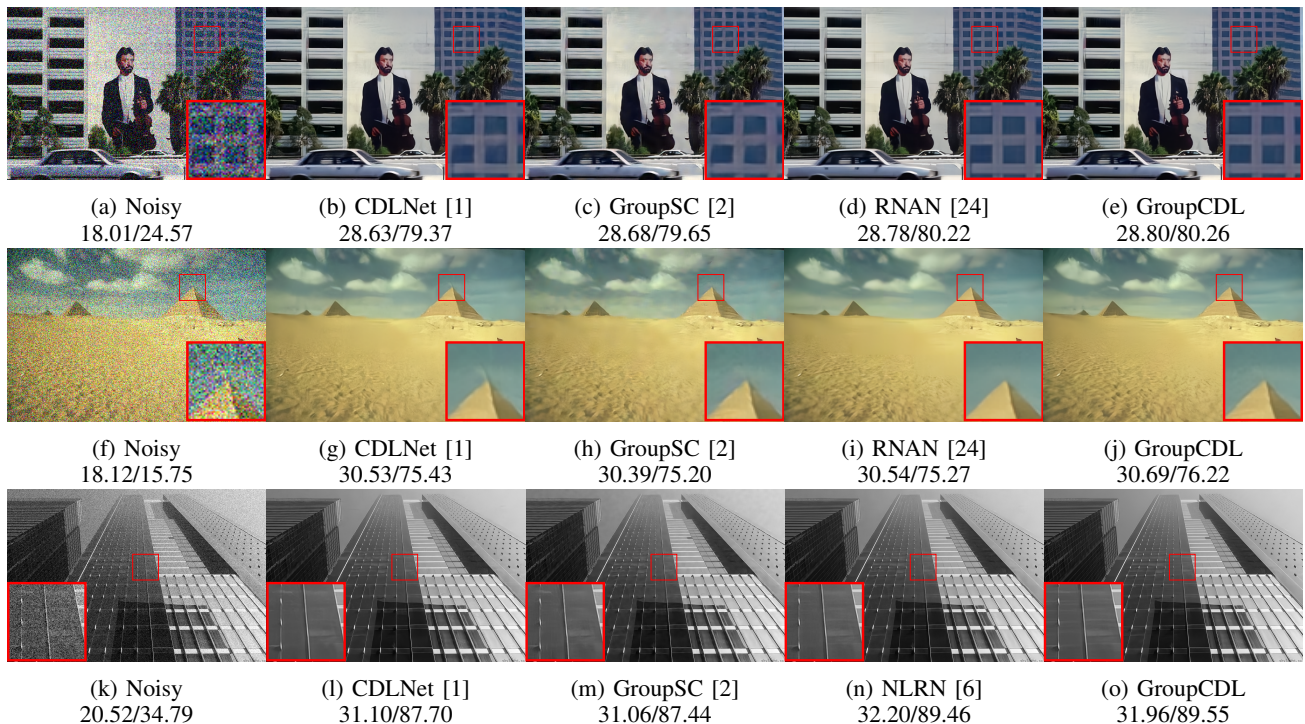

    \centering
    \captionsetup[subfigure]{justification=centering}
    \input{figs/posterman.tex}
    \\
    \input{figs/pyramid.tex}
    \\
    \input{figs/urban.tex}
    \caption{
    Visual comparison of deep denoisers. Top and middle rows:
    color denoisers for $\sigmatrain=\sigmatest=50$. Bottom row: grayscale
    denoisers $\sigmatrain=\sigmatest=25$. PSNR/$100\times$SSIM shown in
    respective captions. 
    }
    \label{fig:colorsingle}
\end{figure*}

Figures \ref{fig:graysingle} and \ref{fig:colorsingle} highlight the qualitative differences between the previously mentioned methods. We observe that the group-sparsity
prior of GroupCDL is instrumental in suppressing unwanted denoising artifacts (present in CDLNet's results),
especially in constant image regions where edges are otherwise hallucinated (see Figure \ref{fig:colorsingle} (g) vs (j)). Note that GroupSC produces low spatial-frequency artifacts throughout the image, likely introduced by independent processing of image patches (see Figure \ref{fig:colorsingle} (c,h,m)). Further, GroupCDL's results appear qualitatively indistinguishable to those of \soa black-box models, at a fraction of the inference time.

\subsection{Noise-Level Generalization}
\begin{figure}
    \centering
    \includegraphics[width=\columnwidth]{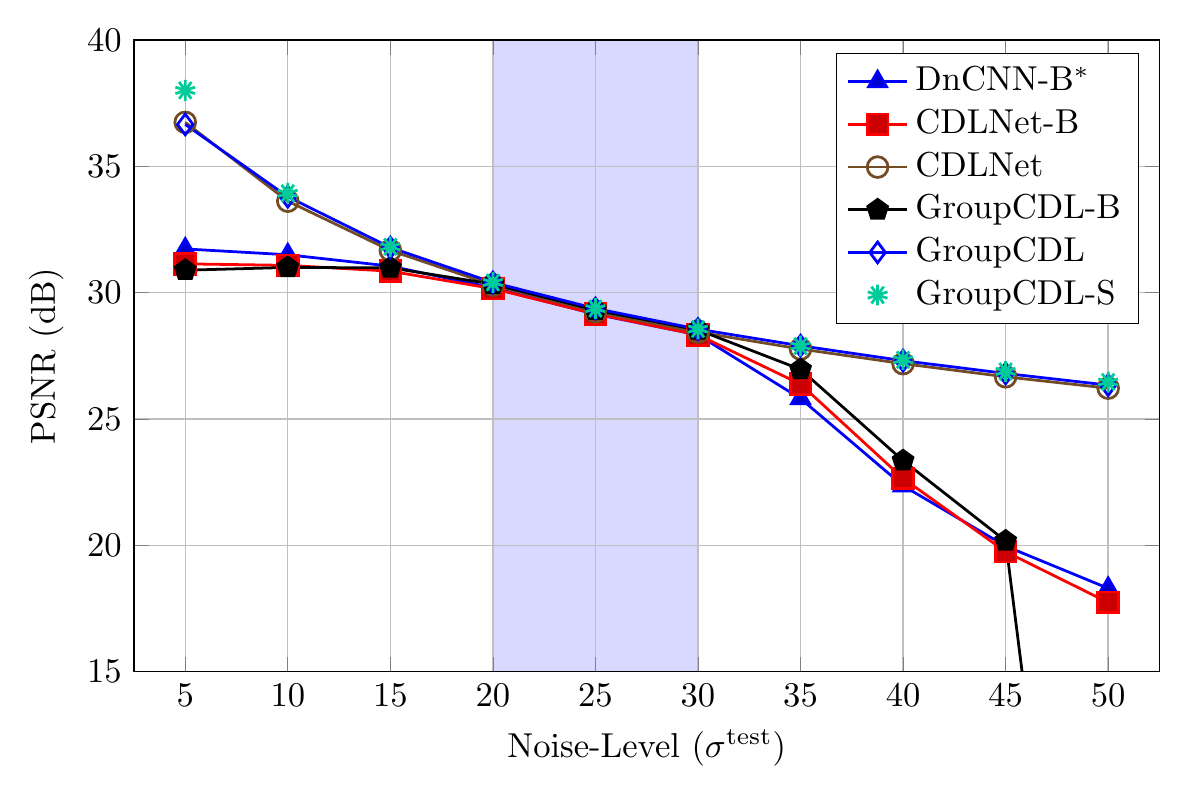}
    \caption{Noise-level generalization of different grayscale denoising
    networks tested on BSD68 \cite{bsd}. GroupCDL-S is trained at $\sigmatest$ 
    for each point on the graph. All other networks are trained on $\sigmatrain
    =[20, 30]$.}
    \label{fig:generalization}
\end{figure}%
\begin{figure}[tbh]
    \centering
    \includegraphics[width=\columnwidth]{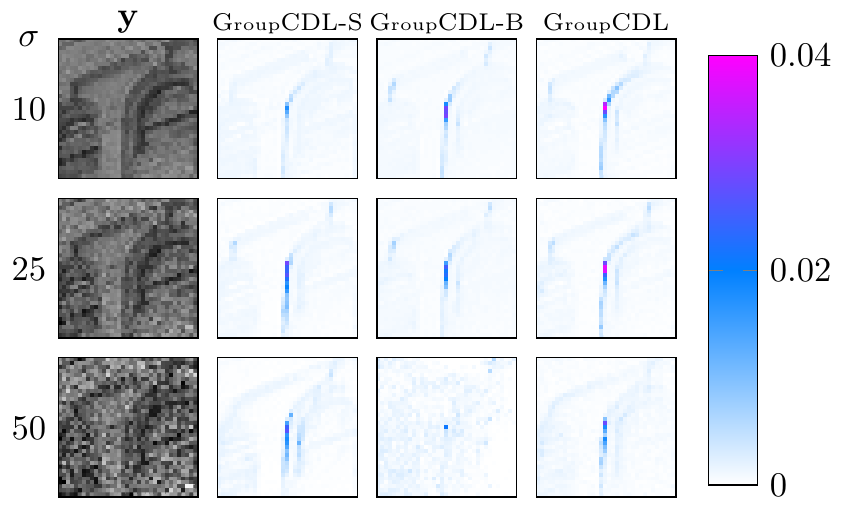}
    \caption{
        Visualization of normalized adjacency $\ADJMAT^{(K)}_{i:}$ for input
        nonlocal window $\y$ at noise-levels $\sigma=10,25,50$. Columns
        GroupCDL-S, GroupCDL-B, and GroupCDL
        visualize the adjacency of models trained under conditions of
        $\sigmatrain=\sigma$, $\sigmatrain \in [20,30]$ without noise-adaptive
        thresholds, and $\sigmatrain \in [20,30]$ with noise-adaptive
        thresholds, respectively. Catastrophic failure of GroupCDL-B model is
        observed for $\sigma > \sigmatrain$. The same noise-realization is used
        across models and noise-levels (with appropriate scaling). 
    }
    \label{fig:adj_gen}
\end{figure}%
In this section, we look at the denoising performance of trained denoising DNNs 
on inference of input images with noise-levels ($\sigmatest$) outside their
training noise-level range ($\sigmatrain$). Figure \ref{fig:generalization}
shows the performance of grayscale image denoisers, trained over the range
$\sigmatrain = [20,30]$. The figure shows, as also noted in \cite{Mohan2020, janjusevicCDLNet2022,
janjusevicGDLNet2022}, black-box DNNs (such as DnCNN \cite{DnCNN}) and dictionary
learning DNNs without noise-adaptive thresholds exhibit a catastrophic failure
on inference above their training noise-level range. A less striking but
analogous failure is seen on inference below the training noise-level,
where a mere plateau in performance is obtained as the denoising problem becomes
easier. 

In addition to the observations noted in \cite{janjusevicCDLNet2022}, Figure
\ref{fig:generalization} shows that the proposed novel group-thresholding
scheme \eqref{eq:GTlearned} is able to obtain near-perfect noise-level
generalization (w.r.t GroupCDL-S performance). This serves as empirical evidence for
the interpretation of the unrolled network as performing some
approximate/accelerated group-sparse BPDN, as the noise-adaptive thresholds
($\btau = \btau_0 + \hat{\sigma}\btau_1$) appear to correspond very well to
their classical counter-parts from which they are derived. 

We further investigate the behavior of the proposed group-thresholding across
noise-levels in Figure \ref{fig:adj_gen}. The figure shows a single input nonlocal
window $\y$ across noise-levels $\sigma$ and the computed adjacency values of
the three types of GroupCDL models (GroupCDL-S, GroupCDL-B, GroupCDL). 

In agreement with the catastrophic failure of the noise-level blind GroupCDL-B
model of Figure \ref{fig:generalization}, the adjacency visualizations of
GroupCDL-B in Figure \ref{fig:adj_gen} show a catastrophic failure in the
similarity computations of the network above $\sigmatrain$, as no structure is
found. We observe a similar pattern in the
GroupCDL models (with noise-adaptive thresholds) to the visualizations of
GroupCDL-S, however, the single noise-level models seem to have more variation
in their computed similarities across noise-levels.
This could suggest room for improvement in GroupCDL by parametrizing the similarity computations
of \eqref{eq:GT} and \eqref{eq:sim} to also be noise-adaptive. However, this subtle
dissimilarity may also be due to the fact that each row-element for the GroupCDL-S
column in Figure \ref{fig:adj_gen} is from a different trained model with
different weights (i.e. with similarities computed in different domains),
whereas the visualizations from the GroupCDL column are all from the same model.

\subsection{Sliding vs. Overlapping Window Nonlocal Processing} \label{sec:exp:slidingwindow}
From Figure \ref{fig:nlss}, it is clear that the OW-NLSS strategy 
(used by NLRN \cite{liu2018non} and GroupSC \cite{lecouat2020nonlocal})
is able to achieve a denoising to speed trade-off by reducing the amount of overlap between windows, i.e.
increasing window-stride ($s_w$). SW-NLSS can also achieve a similar performance-speed trade-off by instead 
reducing the nonlocal window-size ($W$) during inference.
\begin{figure}[thb]
    \centering
    \begin{subfigure}{0.4\textwidth}
        \centering
        \includegraphics[width=\textwidth, page=1]{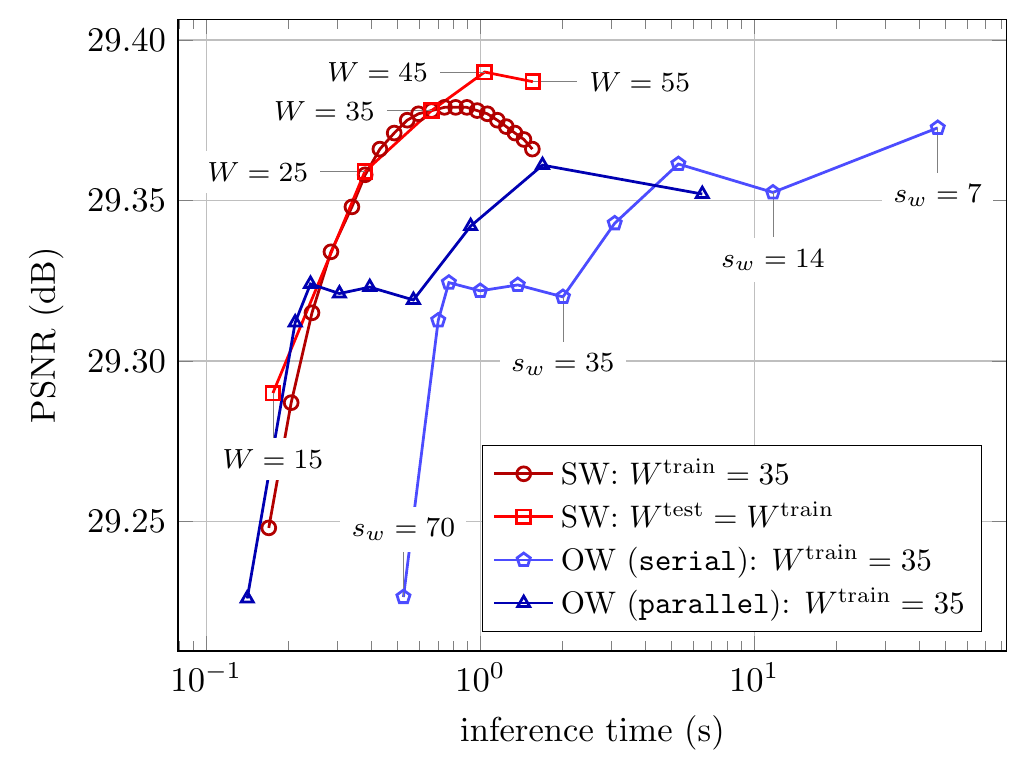}
    \end{subfigure}
    \hfill 
    \begin{subfigure}{0.4\textwidth}
        \centering
        \includegraphics[width=\textwidth, page=2]{figs/graph_owsw.pdf}
    \end{subfigure}
    \caption{
    Inference-time vs. denoising-performance trade-off for GroupCDL. 
    Circle, pentagon, and triangle markers are generated by the same trained GroupCDL model ($\Wtrain = 35$) 
    under different inference strategies (SW-NLSS w/ window-size $W$, OW-NLSS w/ window-stride $s_w$). The square markers are each generated by a GroupCDL with a different training window-size $\Wtrain$. 
    Performance (a) PSNR, (b) $100 \times $SSIM,
    is evaluated on BSD68 \cite{bsd} with $\sigmatrain=\sigmatest=25$. OW-NLSS
    (\texttt{serial,parallel}) curves are generated via processing independent overlapping
    windows either sequentially or all at once, respectively. The same
    noise-realization for the dataset was used across all evaluations plotted.
    Note that $W$ corresponds between SW curve markers vertically (the same inference time), 
    and $s_w$ corresponds between OW curve markers horizontally (the same PSNR/SSIM).
    }
    \label{fig:owsw_curves}
\end{figure}%
\let\myfig\undefined
\newcommand{\myfig}[3]{%
\begin{subfigure}{0.24\linewidth}
    \centering
    \includegraphics[width=\linewidth, page=#2]{#1}
    \caption{#3}
\end{subfigure}%
}

\begin{figure*}[thb]
\centering
\myfig{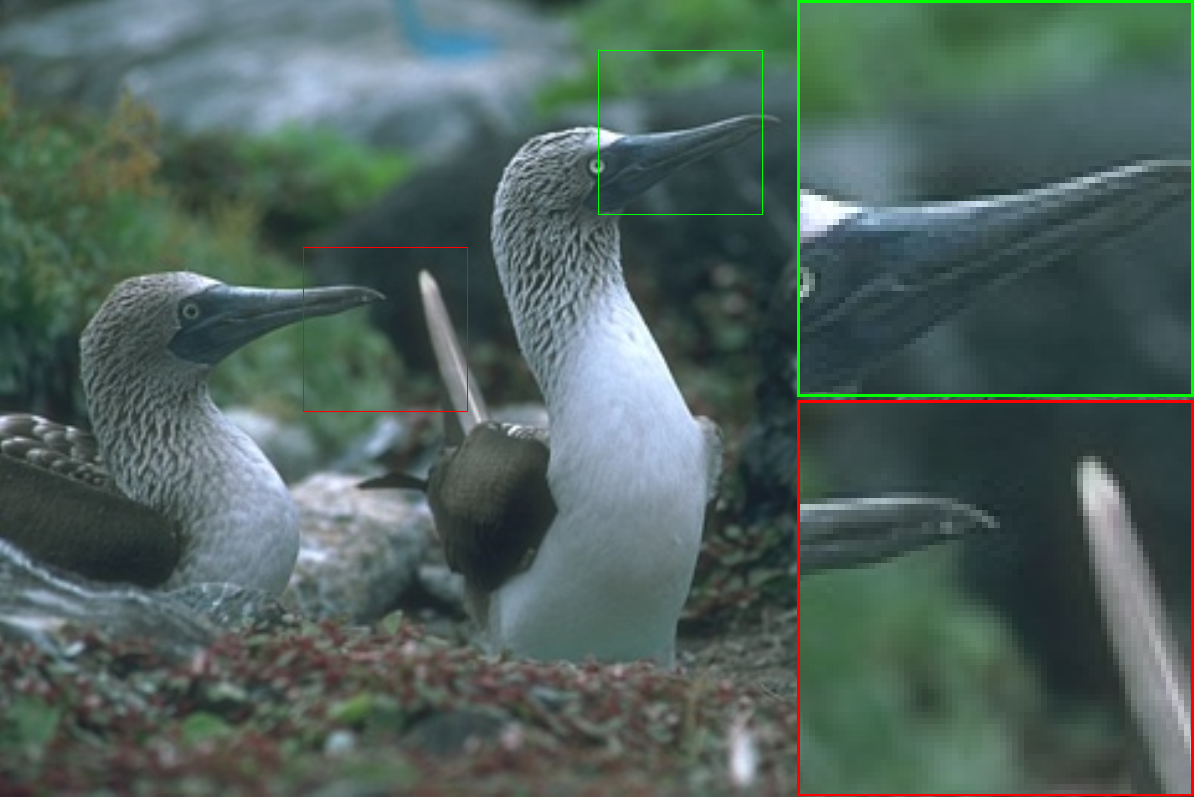}{2}{Noisy}
\myfig{figs/bird_owsw_images.pdf}{3}{$s_w=70$, 29.59 / 82.83 / 0.09}
\myfig{figs/bird_owsw_images.pdf}{5}{$s_w=35$, 29.72 / 83.29 / 0.33}
\myfig{figs/bird_owsw_images.pdf}{6}{$s_w=32$, 29.72 / 83.45 / 0.40}
\\
\myfig{figs/bird_owsw_images.pdf}{1}{Ground Truth}
\myfig{figs/bird_owsw_images.pdf}{9}{$W=15$, 29.57 / 82.92 / 0.08}
\myfig{figs/bird_owsw_images.pdf}{10}{$W=25$, 29.79 / 83.69 / 0.16}
\myfig{figs/bird_owsw_images.pdf}{11}{$W=35$, 29.83 / 83.75 / 0.29}

\caption{
    Comparison of inference-speed/denoising trade-off between OW-NLSS (b-d) and
    SW-NLSS (f-h) processing by a color GroupCDL model
    ($\sigmatrain=\sigmatest=50$, $W^{\mathrm{train}}=35$). PSNR (dB) /
    $100\times$SSIM / GPU inference-time (s) shown in respective captions.
    Zoomed-in regions highlight blocking artifacts exist across shown OW-NLSS
    window-strides ($s_w$), whereas SW-NLSS processing exhibits no blocking
    artifacts across inference window-sizes ($W$). Yellow arrows (b-d) point to
    specific blocking artifact boundaries of OW-NLSS. Orange arrows (f) point
    to edge/texture hallucination artifacts of SW-NLSS with a small windowsize.
    The inclusion of (d) ($s_w=32$), demonstrates that blocking artifacts are
    not merly a result of the effective spatial windowsize ($s_cW=70$) being
    divisible by the overlapping window-stride. The same noise-realization is
    used across all methods (b-d, f-h). 
}
\label{fig:color_owsw}
\end{figure*}

Figure \ref{fig:owsw_curves} plots the denoising performance vs. inference time
trade-off attainable by GroupCDL under the SW-NLSS strategy and the OW-NLSS
strategy. The SW: $\Wtrain =35$, and OW curves show a single GroupCDL model
(trained with a nonlocal window-size $\Wtrain=35$) under SW-NLSS inference with
varying nonlocal window-size $\Wtest$, and OW-NLSS inference with varying
window-stride $s_w$, respectively. Each point in the SW: $\Wtrain = \Wtest$
curve shows the performance of a GroupCDL model trained with a different
non-local window-size and performing SW-NLSS inference with their respective
training window-size. 

First, we observe that SW-NLSS consistently
out-performs OW-NLSS across the PSNR-speed and SSIM-speed trade-offs -- in both
parallel and serial window processing forms of OW-NLSS. This agrees well with the burden-factor analysis in Section \ref{sec:slidingwindow}. The curves further highlight that competitive denoising performance in OW-NLSS inference is predicated on using a small window-stride, in order to compensate for the neglect of dependencies between overlapping window regions by window-averaging.
Second, we observe that increasing the training window-size $\Wtrain$ has
diminishing returns on denoising performance. This is consistent with the
intuition that non-local similarities of natural images are generally located
close to the pixel of interest. 
Lastly, we observe that the OW-NLSS curves of Figure \ref{fig:owsw_curves} are
not monotonically increasing with smaller window-stride, and in-fact have
significant drops in the SSIM curves (Fig. \ref{fig:owsw_curves} (b), blue triangle and pentagon curves). The
source of this behavior is windowing artifacts, highlighted visually in Figure
\ref{fig:color_owsw}. These visualizations show that the denoising-speed
trade-off exhibited by OW-NLSS processing comes with the penalty of unnatural
artifacts in the form of grid-lines corresponding to the spatial pattern of
window overlaps. 
In contrast, the proposed SW-NLSS does not exhibit windowing artifacts. Instead, as the window-size decreases, SW-NLSS processing transitions to fully convolutional CDLNet processing, and 
artifacts associated with FCNNs (such as hallucinated edges) are observed.

\subsection{Ablation Studies} \label{sec:ablation}
In this section we examine the denoising and inference time performance of the
GroupCDL model under different hyperparameters associated with the proposed
group-thresholding operation \eqref{eq:GTlearned}, \eqref{eq:sim}. Table
\ref{tab:ablation:update} shows the effect of the update-period parameter
($\Delta K$), which determines how often we update the adjacency $\ADJMAT^{(k)}$
in the network (every $\Delta K$ layers, see Alg. \ref{alg:GroupCDL}). We observe that decreasing the update
frequency increases denoising performance, with diminishing returns, at cost to
the inference speed of the network.
\begin{table}[ht]
    \centering
    \caption{Effect of update frequency ($\Delta K$). Grayscale denoising performance
    averaged over the NikoSet10 dataset ($\sigmatrain=\sigmatest=25$).}
    \begin{tabular}{c|cc}
        \hline
        $\Delta K$ & PSNR/$100\times$SSIM & time (s)\\
        \hline
        2 & 30.22/86.88 & 4.01 \\
        3 & 30.22/86.87 & 4.72 \\
        5 & 30.21/86.87 & 3.47 \\
        10 & 30.21/86.87 & 3.24 \\
        15 & 30.19/86.79 & 3.18 \\
        \hline
    \end{tabular}
    \label{tab:ablation:update}
\end{table}
\begin{table}[ht]
    \centering
    \caption{Effect of NLSS feature compression. Grayscale denoising performance
    averaged over the NikoSet10 dataset ($\sigmatrain=\sigmatest=25$).}
    \begin{tabular}{ccc|cc}
        \hline
        feature-compression & $M_h$ & \texttt{sim\_fun} & PSNR/$100\times$SSIM & time (s) \\
        \hline
        none & n/a & -distance & 30.19/86.79 & 8.64 \\
        $\bW_{\{\theta, \phi\}}$ & 64 & -distance & 30.21/86.86 & 8.13 \\
        $\bW_{\{\theta, \phi, \alpha, \beta\}}$ & 169 & -distance & 30.21/86.89 & 8.89 \\
        $\bW_{\{\theta, \phi, \alpha, \beta\}}$ & 64 & -distance & 30.21/86.87 & 3.47 \\
        $\bW_{\{\theta, \phi, \alpha, \beta\}}$ & 32 & -distance & 30.20/86.84 & 2.00 \\
        $\bW_{\{\theta, \phi, \alpha, \beta\}}$ & 64 & dot & 30.11/86.55 & 3.43 \\
        \hline
    \end{tabular}
    \label{tab:ablation:nlss}
\end{table}

Table \ref{tab:ablation:nlss} shows the effect of employing learned pixel-wise
transforms in the similarity computation \eqref{eq:sim} ($\bW_\theta$, $\bW_\phi$) and
group-thresholding \eqref{eq:GTlearned} ($\bW_\alpha$, $\bW_\beta$). The table
also shows the effect of employing channel reduction in these transforms ($M_h
<< M = 169$) and a comparison of using the negative distance similarity (as
presented in \eqref{eq:sim}) or, as more commonly used in black-box nonlocal and
transformer DNNs, the dot-product similarity \eqref{eq:dot_sim}. From these
experiments, we observe that the use of pixel-wise transforms increases
denoising performance. Most importantly, we observe only a marginal decrease in
performance for setting the latent similarity channel dimension less than the
latent subband dimension $M_h < M$. This marginal decrease is met by a massive
reduction in GPU inference time, which is predicted well by the channel
reduction ration $M / M_h = 169/64 \approx 8.89/3.47$. This demonstrates one of
the advantages of the proposed group-thresholding operation over black-box
dot-product attention, in that the latent similarity channel dimension $M_h$ is decoupled
from the layer's output channel dimension and can be tuned
to achieve a better trade-off between speed and performance.

\section{Discussion and Conclusion}
GroupCDL adapts the classical, patch-processing based, group-sparsity
prior to convolutional sparse coding (CSC) and applies it to the
direct-parametrization unrolling frame-work of CDLNet
\cite{janjusevicCDLNet2022}.
In contrast to the group-sparsity prior of Lecouat et. al's 
patch-based dictionary learning network (GroupSC) \cite{lecouat2020nonlocal},
which employ's overlapping window processing, we formulate our group-sparsity
prior on convolutional sparse codes, and in doing so naturally arrive at a
sliding-window nonlocal self-similarity consistent with the CSC model. 
As discussed in Section \ref{sec:slidingwindow} and empirically validated in Section
\ref{sec:exp:slidingwindow}, the proposed SW-NLSS enjoys an
improved denoising performance over OW-NLSS by properly accounting for
correlations between neighboring image regions and centering similarity computations on each latent pixel of interest. The sparse array arithmetic
employed at inference time, enables orders of magnitude speed-up for the
proposed design compared to \soa competitor networks (Section
\ref{sec:exp:single}). 

Notably, GroupCDL's performance comes without the use of common deep-learning
operations, such as batch-normalization \cite{Ioffe2015} or residual learning
\cite{DnCNN}, and instead relies on the tools of classical signal processing and
optimization such as basis-pursuit denoising and proximal operators.

The fast inference of GroupCDL is aided by a
novel decoupling of the latent subband dimension ($M$) from the hidden
adjacency/similarity dimension ($M_h$) (see Equation \eqref{eq:GTlearned}).
This allows the computational bottleneck of sparse-matrix dense-vector
multiplication to be tuned without harming the capacity of the latent
representation (Table \ref{tab:ablation:nlss}) -- something which is not
achievable in the dot-product attention employed by black-box nonlocal and
transformer networks (Section \ref{sec:slidingwindow}).

Similar to CDLNet \cite{janjusevicCDLNet2022}, GroupCDL is formulated as a direct
parameterization of the proximal gradient method \eqref{eq:pgm}. We show that this derivation allows for near
perfect generalization outside of its training noise-level range (Fig.
\ref{fig:generalization}), simply by parametrizing its thresholds as an affine
function of the input noise-level, as suggested by the classical BPDN
formulation \eqref{eq:bpdn} and the universal thresholding theorem
\cite{Mallat, janjusevicCDLNet2022}. In contrast, black-box networks \cite{DnCNN} 
are shown to fail catastrophically above the noise-level range and simply
plateau in performance when the noise-level decreases (i.e. the problem becomes
easier). In GroupCDL, generalization is additionally observed in its adjacency matrix (Fig. \ref{fig:adj_gen}).

In future work, we aim to adapt GroupCDL to other imaging modalities. We
believe GroupCDL's speed, performance, interpretability, and robustness are
well suited to tackle large signal reconstruction problems with nonlocal
image-domain artifacts, such compressed sensing magnetic resonance imaging. The
unsupervised learning and demosaicing work of CDLNet
\cite{janjusevicCDLNet2022} may be adapted for GroupCDL to this end. 

\section*{Acknowledgments}
The authors would like to thank NYU HPC for its computing resources and
technical support. The authors are grateful to Che Maria Baez for her
linguistic revisions on a preliminary draft of this manuscript.

\ifCLASSOPTIONcaptionsoff
  \newpage
\fi
\bibliographystyle{IEEEtran}
\bibliography{IEEEabrv,references}

\begin{thebibliography}{10}
\providecommand{\url}[1]{#1}
\csname url@samestyle\endcsname
\providecommand{\newblock}{\relax}
\providecommand{\bibinfo}[2]{#2}
\providecommand{\BIBentrySTDinterwordspacing}{\spaceskip=0pt\relax}
\providecommand{\BIBentryALTinterwordstretchfactor}{4}
\providecommand{\BIBentryALTinterwordspacing}{\spaceskip=\fontdimen2\font plus
\BIBentryALTinterwordstretchfactor\fontdimen3\font minus
  \fontdimen4\font\relax}
\providecommand{\BIBforeignlanguage}[2]{{%
\expandafter\ifx\csname l@#1\endcsname\relax
\typeout{** WARNING: IEEEtran.bst: No hyphenation pattern has been}%
\typeout{** loaded for the language `#1'. Using the pattern for}%
\typeout{** the default language instead.}%
\else
\language=\csname l@#1\endcsname
\fi
#2}}
\providecommand{\BIBdecl}{\relax}
\BIBdecl

\bibitem{janjusevicCDLNet2022}
N.~Janjušević, A.~Khalilian-Gourtani, and Y.~Wang, ``{CDLNet}: Noise-adaptive
  convolutional dictionary learning network for blind denoising and
  demosaicing,'' \emph{IEEE Open Journal of Signal Processing}, vol.~3, pp.
  196--211, 2022.

\bibitem{lecouat2020nonlocal}
B.~Lecouat, J.~Ponce, and J.~Mairal, ``Fully trainable and interpretable
  non-local sparse models for image restoration,'' in \emph{European Conference
  on Computer Vision (ECCV)}, 2020.

\bibitem{janjusevicGDLNet2022}
N.~Janjušević, A.~Khalilian-Gourtani, and Y.~Wang, ``Gabor is enough:
  Interpretable deep denoising with a gabor synthesis dictionary prior,'' in
  \emph{2022 IEEE 14th Image, Video, and Multidimensional Signal Processing
  Workshop (IVMSP)}, 2022, pp. 1--5.

\bibitem{Simon2019}
D.~Simon and M.~Elad, ``Rethinking the {CSC} model for natural images,'' in
  \emph{Advances in Neural Information Processing Systems}, 2019, pp.
  2274--2284.

\bibitem{Scetbon2019DeepKD}
M.~Scetbon, M.~Elad, and P.~Milanfar, ``Deep k-svd denoising,'' \emph{IEEE
  Transactions on Image Processing}, vol.~30, pp. 5944--5955, 2019.

\bibitem{liu2018non}
D.~Liu, B.~Wen, Y.~Fan, C.~C. Loy, and T.~S. Huang, ``Non-local recurrent
  network for image restoration,'' in \emph{Advances in Neural Information
  Processing Systems}, 2018, pp. 1680--1689.

\bibitem{julia}
\BIBentryALTinterwordspacing
J.~Bezanson, A.~Edelman, S.~Karpinski, and V.~B. Shah, ``Julia: A fresh
  approach to numerical computing,'' \emph{SIAM Review}, vol.~59, no.~1, pp.
  65--98, 2017. [Online]. Available: \url{https://doi.org/10.1137/141000671}
\BIBentrySTDinterwordspacing

\bibitem{Mallat}
S.~Mallat, \emph{A Wavelet Tour of Signal Processing: The Sparse Way}.\hskip
  1em plus 0.5em minus 0.4em\relax Elsevier Science, 2008.

\bibitem{Beck2009}
A.~Beck and M.~Teboulle, ``A fast iterative shrinkage-thresholding algorithm
  for linear inverse problems,'' \emph{SIAM Journal on Imaging Sciences},
  vol.~2, pp. 183--202, 01 2009.

\bibitem{mairal2009non}
J.~Mairal, F.~Bach, J.~Ponce, G.~Sapiro, and A.~Zisserman, ``Non-local sparse
  models for image restoration,'' in \emph{Proceedings of 12th IEEE
  International Conference on Computer Vision (ICCV)}, 2009, pp. 2272--2279.

\bibitem{mairal2009online}
J.~Mairal, F.~Bach, J.~Ponce, and G.~Sapiro, ``Online dictionary learning for
  sparse coding,'' in \emph{Proceedings of the 26th International Conference on
  Machine Learning (ICML)}, 2009, pp. 689--696.

\bibitem{ongie2020deep}
G.~Ongie, A.~Jalal, C.~A. Metzler, R.~G. Baraniuk, A.~G. Dimakis, and
  R.~Willett, ``Deep learning techniques for inverse problems in imaging,''
  \emph{IEEE Journal on Selected Areas in Information Theory}, vol.~1, no.~1,
  pp. 39--56, 2020.

\bibitem{Gilton2019}
D.~Gilton, G.~Ongie, and R.~Willett, ``Neumann networks for linear inverse
  problems in imaging,'' \emph{IEEE Transactions on Computational Imaging},
  vol.~6, pp. 328--343, 2019.

\bibitem{deqWillet2021}
------, ``Deep equilibrium architectures for inverse problems in imaging,''
  \emph{IEEE Transactions on Computational Imaging}, vol.~7, pp. 1123--1133,
  2021.

\bibitem{unet}
O.~Ronneberger, P.~Fischer, and T.~Brox, ``{U-Net}: Convolutional networks for
  biomedical image segmentation,'' \emph{Medical Image Computing and
  Computer-Assisted Intervention}, p. 234–241, 2015.

\bibitem{he2016deep}
K.~He, X.~Zhang, S.~Ren, and J.~Sun, ``Deep residual learning for image
  recognition,'' in \emph{Proceedings of the IEEE conference on computer vision
  and pattern recognition}, 2016, pp. 770--778.

\bibitem{Zheng_2021_CVPR}
H.~Zheng, H.~Yong, and L.~Zhang, ``Deep convolutional dictionary learning for
  image denoising,'' in \emph{Proceedings of the IEEE/CVF Conference on
  Computer Vision and Pattern Recognition (CVPR)}, June 2021, pp. 630--641.

\bibitem{Sreter2018}
H.~Sreter and R.~Giryes, ``Learned convolutional sparse coding,'' in
  \emph{Proceedings of IEEE International Conference on Acoustics, Speech and
  Signal Processing (ICASSP)}, 2018, pp. 2191--2195.

\bibitem{Gregor2010}
K.~Gregor and Y.~LeCun, ``Learning fast approximations of sparse coding,'' in
  \emph{Proceedings of the 27th International Conference on Machine Learning
  (ICML)}, 2010, pp. 399--406.

\bibitem{child2019sparsetransformer}
R.~Child, S.~Gray, A.~Radford, and I.~Sutskever, ``Generating long sequences
  with sparse transformers,'' \emph{URL
  https://openai.com/blog/sparse-transformers}, 2019.

\bibitem{dao2022flashattention}
T.~Dao, D.~Y. Fu, S.~Ermon, A.~Rudra, and C.~R{\'e}, ``Flash{A}ttention: Fast
  and memory-efficient exact attention with {IO}-awareness,'' in \emph{Advances
  in Neural Information Processing Systems}, 2022.

\bibitem{Mei_2021_CVPR}
Y.~Mei, Y.~Fan, and Y.~Zhou, ``Image super-resolution with non-local sparse
  attention,'' in \emph{Proceedings of the IEEE/CVF Conference on Computer
  Vision and Pattern Recognition (CVPR)}, June 2021, pp. 3517--3526.

\bibitem{Khashabi2014}
D.~Khashabi, S.~Nowozin, J.~Jancsary, and A.~W. Fitzgibbon, ``Joint demosaicing
  and denoising via learned nonparametric random fields,'' \emph{IEEE
  Transactions on Image Processing}, vol.~23, pp. 4968--4981, 2014.

\bibitem{zhang2019residual}
Y.~Zhang, K.~Li, B.~Zhong, and Y.~Fu, ``Residual non-local attention networks
  for image restoration,'' in \emph{International Conference on Learning
  Representations}, 2019.

\bibitem{bsd}
D.~Martin, C.~Fowlkes, D.~Tal, and J.~Malik, ``A database of human segmented
  natural images and its application to evaluating segmentation algorithms and
  measuring ecological statistics,'' in \emph{Proceedings of 8th IEEE
  International Conference on Computer Vision (ICCV)}, vol.~2, 2001, pp.
  416--423.

\bibitem{adam}
D.~P. Kingma and J.~Ba, ``Adam: A method for stochastic optimization,'' in
  \emph{International Conference on Learning Representations (ICLR)}, 2015.

\bibitem{Urban100}
J.-B. Huang, A.~Singh, and N.~Ahuja, ``Single image super-resolution from
  transformed self-exemplars,'' in \emph{Proceedings of the IEEE Conference on
  Computer Vision and Pattern Recognition (CVPR)}, June 2015.

\bibitem{glorot}
\BIBentryALTinterwordspacing
X.~Glorot and Y.~Bengio, ``Understanding the difficulty of training deep
  feedforward neural networks,'' in \emph{Proceedings of the Thirteenth
  International Conference on Artificial Intelligence and Statistics}, ser.
  Proceedings of Machine Learning Research, Y.~W. Teh and M.~Titterington,
  Eds., vol.~9.\hskip 1em plus 0.5em minus 0.4em\relax Chia Laguna Resort,
  Sardinia, Italy: PMLR, 13--15 May 2010, pp. 249--256. [Online]. Available:
  \url{https://proceedings.mlr.press/v9/glorot10a.html}
\BIBentrySTDinterwordspacing

\bibitem{bm3d}
K.~Dabov, A.~Foi, V.~Katkovnik, and K.~Egiazarian, ``Image denoising by sparse
  3-{D} transform-domain collaborative filtering,'' \emph{IEEE Transactions on
  Image Processing}, vol.~16, no.~8, pp. 2080--2095, 2007.

\bibitem{bm3d-gpu}
D.~Honzátko and M.~Kruliš, ``Accelerating block-matching and 3d filtering
  method for image denoising on gpus,'' 11 2017.

\bibitem{DnCNN}
K.~Zhang, W.~Zuo, Y.~Chen, D.~Meng, and L.~Zhang, ``Beyond a gaussian denoiser:
  Residual learning of deep {CNN} for image denoising,'' \emph{IEEE
  Transactions on Image Processing}, vol.~26, no.~7, p. 3142–3155, 2017.

\bibitem{Valsesia2020}
D.~Valsesia, G.~Fracastoro, and E.~Magli, ``Deep graph-convolutional image
  denoising,'' \emph{IEEE Transactions on Image Processing}, vol.~29, pp.
  8226--8237, 2020.

\bibitem{Mohan2020}
S.~Mohan, Z.~Kadkhodaie, E.~P. Simoncelli, and C.~Fernandez-Granda, ``Robust
  and interpretable blind image denoising via bias-free convolutional neural
  networks,'' in \emph{International Conference on Learning Representations
  (ICLR)}, 2020.

\bibitem{Ioffe2015}
S.~Ioffe and C.~Szegedy, ``Batch normalization: Accelerating deep network
  training by reducing internal covariate shift,'' in \emph{International
  conference on machine learning}.\hskip 1em plus 0.5em minus 0.4em\relax PMLR,
  2015, pp. 448--456.

\end{thebibliography}

\end{document}